# Testing the $H_2O_2$-$H_2O$ Hypothesis for Life on Mars with the TEGA instrument on the Phoenix Lander

Dirk Schulze-Makuch[1], Carol Turse[2], Joop M. Houtkooper[3], and Chris McKay[4]

[1]School of Earth and Environmental Sciences, Washington State University, Pullman, WA 99164, Tel: 1-509-335-1180, Fax: 1-509-335-7816, email dirksm@wsu.edu
[2]School of Earth and Environmental Sciences, Washington State University, Pullman, WA 99164, email: carol@turse.org
[3] Center of Psychobiology and Behavioral Medicine, Justus-Liebig University of Giessen, Otto-Behaghel-Strasse 10F, D-35394 Giessen, Germany, email: joophoutkooper@gmail.com
[4]NASA Ames, M.S. 245-3, Moffett Field, CA 94035, Tel: 650-604-6864, Fax: 650-604-6779, email cmckay@mail.arc.nasa.gov

**Abstract**

Since Viking has conducted its life detection experiments on Mars, many missions have enhanced our knowledge about the environmental conditions on the Red Planet. However, the Martian surface chemistry and the Viking lander results remain puzzling. Non-biological explanations that favor a strong inorganic oxidant are currently favored (e.g., Mancinelli, 1989; Quinn and Zent, 1999; Klein, 1999, Yen et al., 2000), but problems remain regarding the life time, source, and abundance of that oxidant to account for the Viking observations (Zent and McKay, 1994). Alternatively, a hypothesis favoring the biological origin of a strong oxidizer has recently been advanced (Houtkooper and Schulze-Makuch, 2007). Here, we report about laboratory experiments that simulate the experiments to be conducted by the Thermal and Evolved Gas Analyzer (TEGA) instrument of the Phoenix lander, which is to descend on Mars in May 2008. Our experiments provide a baseline for an unbiased test for chemical versus biological responses, which can be applied at the time the Phoenix Lander transmits its first results from the Martian surface.

**Background**

The Viking mission left open many questions, particularly the nature of the oxidant in the Martian soil remains enigmatic (Zent and McKay, 1994; Quinn and Zent, 1999; Yen et al., 2000; Mancinelli, 1989; Klein, 1999; Benner et al., 2000; Horowith et al., 2007; Houtkooper and Schulze-Makuch, 2007). The common explanation is that all the organic material near the surface was oxidized by $H_2O_2$ and other strong oxidizing compounds. Based on the reactivity of the surface measured by the Viking Gas Exchange experiment (GEx), the amount of $H_2O_2$ on the Martian surface was estimated to be between 1 ppm (Zent and McKay, 1994) and 250 ppm (Mancinelli, 1989). Yet, photochemical processes generate $H_2O_2$ in the atmosphere at a much lower rate in the parts per billion range. Atmospheric $H_2O_2$ abundances vary between 20 and 40 ppb by volume over the planet (Encrenaz et al. 2004), which appears to be a maximum concentration occurring during favorable weather conditions (Atreya and Gu, 1994). Instead of $H_2O_2$, superoxide ions were suggested as well (Yen et al., 2000). However, the laboratory data presented in the literature deal with bulk, reagent-grade superoxide materials and it is difficult to imagine any existing natural mechanisms on Mars that would produce such bulk superoxides. Nussinov et al. (1978) argued that oxygen gas physically trapped in soil micropores might be responsible for the Viking observations. Another promising hypothesis was advanced by Quinn and Zent (1994) suggesting that hydrogen peroxide chemisorbed on titanium dioxide may be responsible for the chemical reactivity seen in the Viking life detection experiments. Possible oxidant reactions and their environmental problems on Mars were pointed out by Zent and McKay (1994), who concluded that none of the hypotheses presented in the literature is free of serious objections, many having to do with the instability of putative oxidants in the presence of heat, light, or atmospheric carbon dioxide. Or, the suggested hypotheses would require elaborate formation mechanisms for which there is no evidence. However, they also rejected the biological explanation, but rather suggesting that the results obtained by Viking could be best explained by some kind of heterogeneous surface chemistry, yielding one or more types of oxidizing surfaces on the Martian regolith particles.

Alternatively, Levin and Straat (1981) and Levin (2007) argued for a biological explanation, but struggled to explain (1) the evolution of $O_2$ upon wetting the soil, (2) the apparent absence of organic molecules in the soil, and (3) the weakly positive results of the single control test in the Pyrolytic Release experiment. Using a different approach, Houtkooper and Schulze-Makuch (2007) suggested that putative Martian organisms might employ a novel biochemistry; in particular utilize a water-hydrogen peroxide ($H_2O$-$H_2O_2$) mixture rather than water as an intracellular liquid. This adaptation would have the particular advantages of providing a low freezing point, a source of oxygen, and hygroscopicity in the Martian environment, allowing organisms to scavenge water molecules directly from the atmosphere, and address many of the puzzling Viking findings. $H_2O_2$-$H_2O$ solutions are mostly known as disinfectants and sterilizing agents on Earth, but some microbial organisms produce hydrogen peroxide (e.g., certain *Streptococcus* and *Lactobacillus sp.*; Eschenbach et al., 1989), while other microbes utilize $H_2O_2$ (e.g., *Neisseria sicca*, *Haemophilus segnis;* Ryan and Kleinberg, 1995). Sensitivity to $H_2O_2$ varies drastically (Anders et al., 1970; Alcorn et al., 1994; Stewart et al., 2000). Reported microbial survival rates range from greater than 80 % to less than 0.001 % after exposure to 30 mM hydrogen peroxide (Alcorn et al., 1994) and at least one organism, the microbe *Acetobacter peroxidans*, uses $H_2O_2$ in its metabolism (overall reaction $H_2O_2(aq) + H_2(aq) \leftrightarrow 2H_2O$; Tanenbaum, 1956).

The NASA Mars Phoenix Mission, which is currently on its way to Mars, provides the unique possibility to test various hypotheses to explain the Viking results. The Phoenix lander includes the TEGA instrument, a combination high-temperature furnace and mass spectrometer that will be used to analyze Martian ice and soil samples. Once a sample is successfully received and sealed in the TEGA oven, temperature is slowly increased at a constant rate up to 1000°C, and the power required for heating is carefully and continuously monitored. This process, called scanning calorimetry, shows the transitions from solid to liquid to gas of the different materials in the sample.

**Material and Methods**

Differential scanning calorimetry was used by us to analyze phase transitions and thermodynamic properties of the oxidant compounds investigated. Two thermal cells were employed, one cell holding the reference capsule, the other the sample. A computer control system measured the amount of heat required to increase the temperature of each cell. If the temperature in one cell was not rising as fast as the temperature in the other cell, the instrument sent more energy (heat) to the heating coils in that cell to maintain the same temperature in each cell. The computer then recorded the difference between the energy requirements for each cell. The resulting graph (thermogram) of temperature versus energy difference between the two cells displayed a peak whenever a phase transition occurred. The area under a positive peak (peak area) represents the energy required for the transition (enthalpy of the reaction, $\Delta H$), thus any positive peaks are representative of endothermic reactions, while negative peaks are representative of exothermic reactions. The onset of a peak usually corresponds to the melting or evaporation temperature of a tested substance. If the weight of the sample is known, then the Differential Scanning Calorimeter (DSC) can calculate the energy required per gram of sample (J/g) for the transition. In an exothermic process, less heat would be required by the sample than by the reference cell to keep a steady change in temperature. In this case, the resulting peak on the thermogram is a negative peak. The heat of the phase change can be adsorbed or released depending on the change in specific heat characteristics of each phase.

Sample compounds investigated included millipure water, 17.5% and 35 % hydrogen peroxide solution, 99.9% pure $Fe_2O_3$, 99.9% pure $TiO_2$, tetrasodium pyrophosphate ($Na_4P_2O_7$), phenacetin ($C_{10}H_{13}NO_2$), quartz sand, JSC-1 Martian regolith simulant soil, and combinations thereof. $Na_4P_2O_7$ and phenacetin are stabilizers of $H_2O_2$ (Fig. 1) and were included in the test set, because if the Martian $H_2O_2$ would mostly be of biological nature, a chemical stabilizer has to be invoked to control the reactivity of the hydrogen peroxide. Other potential chemical stabilizers for $H_2O_2$ solutions include sodium silicate ($Na_2SiO_3$), poly($\alpha$)hydroxyacrylic acid, phytate, citrate, and malonate (Charron et al., 2006; Watts et al., 2007). Tetrasodium pyrophosphate was chosen here due to its simplicity, efficacy (e.g., common use in commercial applications) and its similarity to ATP, phenacetin due to its demonstrated long-term (> 3 months) effectiveness to keep hydrogen peroxide stable (Madanská et al., 2004). The Martian regolith simulant is the <1mm fraction of weathered volcanic ash from Pu'u Nene, a cinder cone on the island of Hawaii, and was provided by the Johnson Space Center. The scanning rate of the DSC was set at 10°C per minute and the sample amount used was about 20 mg to simulate the thermograms that will be obtained by the TEGA instrument on Mars. Samples were put in standard 20 μl aluminum sample pans with covers and sealed using the standard sample pan crimper press. Each pan was fully loaded and weighted before and after the addition of the sample. The sample preparation was completed in less than 5 minutes, since it was observed that the hydrogen peroxide started to decompose under atmospheric conditions within 40 min affecting some of the thermograms (Figs. not shown). An indium standard was used to calibrate the DSC and a baseline run using an empty aluminum pan was conducted prior to each sample run. The DSC was programmed to automatically subtract each baseline run from the sample run.

**Results**

The thermograms of the various compounds and solutions are shown in Fig. 2. The millipure water revealed a behavior characteristic of the phase transition from liquid water to water vapor with a peak at about 110°C. Any differences observed in peaks, ranges, and energies in duplicate runs (Fig. 2a-f) were a function of the total amount of solutes used in the DSC and due to instrument variation. The 17.5 % hydrogen peroxide solution revealed a very similar behavior to pure water, but also exhibited a small negative peak at 123°C (Fig. 2b). The negative peak appeared when almost all of the water had evaporated and represents the heat given off by the $H_2O_2$ as it decomposed exothermally. This negative peak was more pronounced when using a 35 % hydrogen peroxide solution (Fig. 2c). The area under the large positive (endothermic) peak of the 35 % hydrogen peroxide solution is 1.096 kJ/g +/- 0.31 kJ/g. The characteristic thermogram of a 35 % $H_2O_2$ solution was clearly identifiable at concentrations down to 50 ppm or about 1ng (not shown), which is within the range of hydrogen peroxide concentrations to be expected on the Martian surface[5]. A solution of 35% hydrogen peroxide with tetra sodium pyrophosphate at a ratio of 9:1 revealed a very different pattern (Fig. 2d). It produced a strong exothermic response at about 80°C, which was likely due to the hydrolysis of the pyrophosphate ion, and a large endothermic peak close to 100°C. The stabilizing effect of the pyrophosphate must have ceased after all the pyrophosphate ions had been hydrolyzed. Notice that the overall energy required for the phase transition greatly increased compared to millipure water and the hydrogen peroxide solutions (scale on y-axis of Fig. 2d). The thermogram of the chemical stabilizer phenacetin revealed a characteristic single peak at about 132°C (Fig. 2e), which was also observed in the 35 % hydrogen peroxide solution when phenacetin was added at a 9:1 mass ratio of 35% $H_2O_2$ to phenacetin (Fig. 2f). The phenacetin peak was observed at a value of 131°C in the hydrogen peroxide solution, while the main peak was at 108°C. The thermogram displayed again a small negative peak at 118°C characteristic for the exothermic decomposition of hydrogen peroxide.

In the next set of sampling runs various soil media and metal oxides were exposed to the previously tested solutions (Fig. 3a-d). Quartz sand displayed a characteristic peak at about 44°C, which appeared in JSC-1 Mars simulant soil more pronounced. The second, larger peak reflects the evaporation of water. The addition of the chemical stabilizers $Na_4P_2O_7$ and phenacetin shifted the peaks toward higher temperatures. In addition, the phenacetin peak was clearly identifiable in all media when it was used at a mass fraction of 3 % (Fig 3a-d). However, when the mass ratio of the phenacetin spiked hydrogen peroxide solution was lowered to 10 % of the mass fraction of the soil (phenacetin ~ 1 % of the total mass of the soil), the signature of the phenacetin was too small to be unambiguously identified in the JSC-1 soil and the metal oxides tested (Fig. not shown). The negative peak of the decomposition of hydrogen peroxide is not discernable when using soil media and metal oxides. When using the JSC-1 Martian simulant soil, which contains many metal oxides, a larger amount of endothermic energy was needed to evaporate the water. The peak energy was reached at significantly higher temperatures compared to quartz sand (Fig. 3a,b). The same pattern is revealed when using titanium- and iron oxides as a medium (Fig. 3c and 3d, respectively). However, the second, larger peak varied much more for the different solutions within metal oxides as a medium. For example, when using $TiO_2$ as a medium, a plateau is displayed for water and most hydrogen peroxide solutions between the two peaks. A prominent peak, however, appeared if phenacetin was added to the solutions and if the tested medium contained metal oxides (JSC-1 soil and metal oxides). In the metal-oxide containing media the phenacetin peak at about 132°C was not as strong as in quartz sand, but still discernable at the concentrations tested.

**Implications and Conclusions**

The thermograms for the JSC-1 simulant soil and the metal oxides are very similar indicating that the thermogram of JSC-1 soil is dominated by the response of its metal oxide composition. The thermograms are very sensitive to moisture and allow an easy detection of water. The exothermic decomposition of $H_2O_2$ is detectable down to concentrations of at least 50 ppm, but more difficult to detect within soil media. Further compounding the difficulty of $H_2O_2$ detection is that laboratory runs under atmospheric conditions indicated that $H_2O_2$ and its characteristic signature decays. This will also be the case under Martian atmospheric conditions. The Phoenix lander is equipped with a soil sampler, however, which should allow in-situ analysis before the oxidant in the soil is degraded too much. Our testing indicates that $H_2O_2$ would be easier to detect by TEGA within an ice sample than in a soil sample.

The addition of a chemical stabilizer to a hydrogen peroxide solution can be identified in the thermogram. More endothermic energy is required during the heating process shifting the peak energy towards higher temperatures. In addition, phenacetin displays a characteristic peak at about 132°C. The detection of the chemical stabilizer is more challenging at lower concentrations within a soil matrix. The phenacetin was clearly identifiable in all four tested media at a tested mass fraction of about 3 % (Fig 3a-d), but the peak was not significant at a tested mass fraction of about 1 %. Concentrations of $Na_4P_2O_7$ have to be even higher in concentration to be clearly discernable in the thermograms.

If we entertain the $H_2O_2$-$H_2O$ hypothesis of Martian life (Houtkooper and Schulze-Makuch, 2007), then an organic stabilizer such as phenacetin could be understood as a more sophisticated evolutionary adaptation of life to Martian conditions than an inorganic stabilizer such as $Na_4P_2O_7$. Both, however, would serve as a biomarker for possible life on Mars. The advantage of phenacetin is that its biosignature can be easier detected in thermograms and also that there is no plausible way of an inorganic production of this organic compound. Alternatively, the $H_2O_2$ in the Martian soil may be due to inorganic processes (Horowith et al., 2007), in which case no stabilizer would be present. Thus, the detection of a sufficient amount of $H_2O_2$ by itself would not provide evidence for the $H_2O_2$-$H_2O$ hypothesis. On the other hand, a measured distinct change in isotope fractionation ratios of $^{13}C/^{12}C$ and $^{18}O/^{16}O$ at temperatures at which organics decompose would be supportive of a biological explanation.

Obviously, the idea that the Viking lander observations, which implied a strong oxidizing agent for the observed reactions (Klein, 1999), could be caused by biology are highly speculative, but so is nearly any conjecture in astrobiology. The main distinction here is that the biological and the chemical hypotheses presented are testable, which is rare in this field of study. The experiments reported on were conducted to simulate the analyses of the TEGA instrument on Mars. Thus, a heating rate of $10^oC$ per minute and the same sampling volume was used. However, even if the heating rate would be reduced, our work suggests that the thermograms would only be smoothened, but that the general trends and shapes of the thermograms would remain the same. The anticipated results to be obtained from the TEGA instrument of the Phoenix lander will thus provide an unbiased test of the nature of the oxidant on Mars and also be helpful in the interpretation of the results of the Sample Analysis at Mars Instrument Suite (SAM) of the Mars Science Laboratory (MSL) mission.

**Fig. Captions**

Figure 1. Structural formula of chemical stabilizers of hydrogen peroxide, a. sodium pyrophosphate, b. phenacetin

Figure 2. Thermograms of various solutions. A. Millipure water with a peak value of 114°C (+/- 3°C). The peak ranged from 80°C (+/- 1°C) to 117°C (+/- 3°C) and the peak area was 1403 J/g (+/- 105 J/g). B. 17.5% hydrogen peroxide solution with a peak value of 117°C (+/- 3°C). The peak area was 1270 J/g (+/- 0.2 J/g) with the peak ranging from 68°C (+/- 4) to 123°C (+/- 2°C). The solution also exhibited a small negative peak at 123°C (+/- 1°C) with an area of –7 J/g (+/- 0.3°C). C. 35% $H_2O_2$ solution with a peak value of 101°C (+/-1°C). The peak ranged from 52°C (+/-1) to 117°C +/-1). The negative peak is at 117°C (+/-1°C) with an area of -27J/g (+/- 5J/g). D. 35% hydrogen peroxide solution with sodium pyrophosphate at a ratio of 9 to 1. The thermogram revealed a large exothermic peak at about 86°C with an area of –103 J/g (+/- 71 J/g). The main peak had a value of 97°C (+/- 3°C), a range from 87°C (+/- 0.5°C) to 107°C (+/- 6°C) and an area of 874 J/g (+/- 570 J/g). E. Phenacetin revealed a single peak at 133°C (+/- 1°C) with an area under the curve of 173 J/g (+/- 8 J/g). F. 35 % hydrogen peroxide solution with 10 % phenacetin. The peak was at 108°C (+/- 1°C) with a range of 61°C (+/- 3°C) to 118°C (+/- 1°C) and an area under the curve of 785 J/g (+/- 37 J/g). The phenacetin peak appears as a minor peak at a value of 131°C (+/- 1°C) with an area of 4 J/g (+/- 1 J/g). The solution also displayed again a small negative peak at 118°C (+/- 1°C) characteristic for the exothermic reaction of $H_2O_2$ with an area under the curve of –8 J/g (+/- 1 J/g).

Figure 3. Thermograms with various media and solutions. A. Quartz sand. The first peaks appear at a value of 44°C (+/- 4°C) with a range of 35°C (+/- 6°C) to 50°C (+/- 9°C) and a peak area of 8 J/g (+/- 1 J/g). The second, larger peaks have a value of 77°C (+/- 5°C) with a range from 45°C (+/- 1°C) to 91°C (+/- 6°C) and a peak area of 117 J/g (+/- 21 J/g). A sharp phenacetin peak is observed at 132°C for the hydrogen peroxide solution with 10 % phenacetin. B. JSC-1 Mars stimulant soil. The first peaks appear at a value of 41°C (+/- 1°C) with a range from 35°C (+/- 1°C) to 46°C (+/- 1°C) and a peak area of 9 J/g (+/- 2 J/g). The second, larger peaks average at a value of 92°C (+/- 2°C) with a range from 55°C (+/- 6°C) to 120°C (+/- 8°C) and a peak area of 224 J/g (+/- 47 J/g). The phenacetine peak is again observed at a value of 132°C, but smaller in magnitude. C. Titanium (IV) oxide. The first peaks have a value of 42°C (+/- 1°C) with a range from 35°C (+/- 4°C) to 46°C (+/- 1°C) and a peak area of 6 J/g (+/- 3 J/g). The second, larger peaks show much variation with a range from 62°C (+/- 16°C) to 90°C (+/- 9°C) and an average peak area of 69 J/g (+/- 41 J/g). The phenacetin peak is observed again at a value of 127°C. D. Iron (III) oxide. The first peaks have a value of 41°C (+/- 3°C) with a range from 37°C (+/- 4°C) to 47°C (+/- 3°C) and a peak area of 7 J/g (+/- 4 J/g). The second, larger peaks vary largely with a range from 49°C (+/- 6°C) to 85°C (+/- 11°C) and an average peak area of 146 J/g (+/- 49 J/g). The phenacetin peak is observed at a value of 130°C.

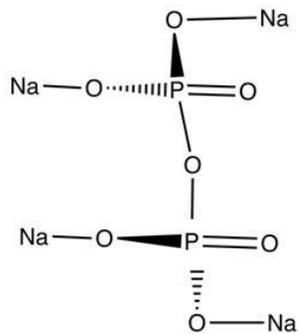

a) Sodium Pyrophosphate

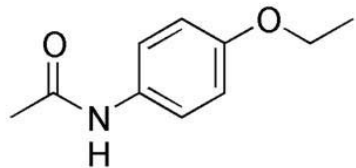

b) Phenacetin

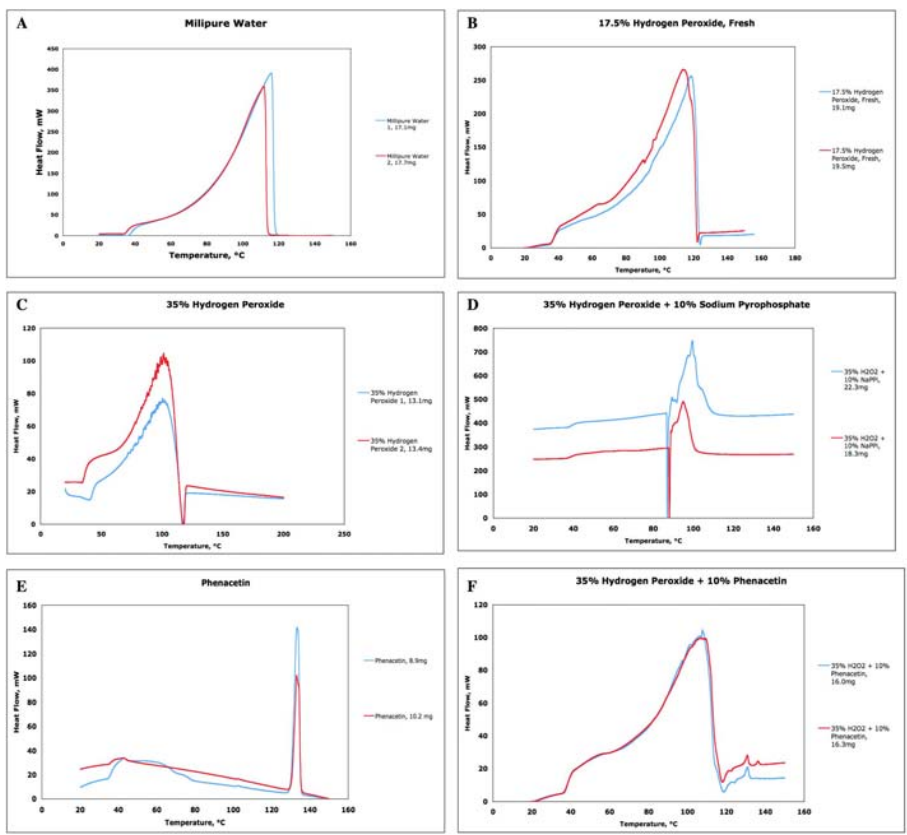

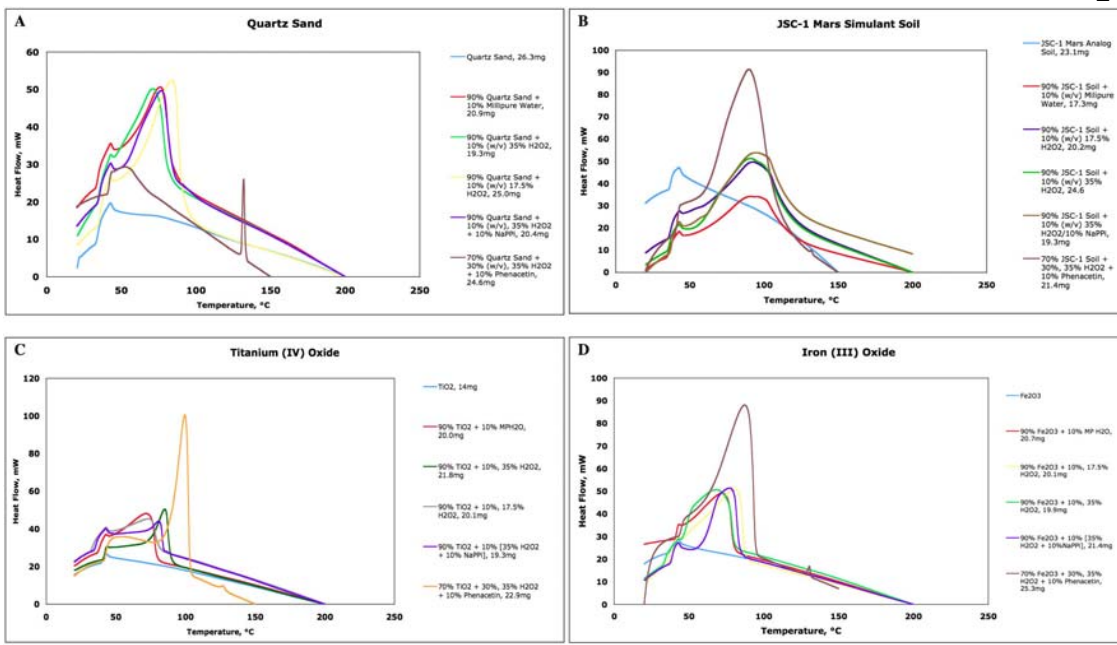